\documentclass[12pt]{article}
\usepackage{epsf}
\def\be{\begin{equation}}
\def\ee{\end{equation}}
\def\bea{\begin{eqnarray}}
\def\eea{\end{eqnarray}}

\begin{document}
\begin{titlepage}
\begin{center}
{\Large \bf William I. Fine Theoretical Physics Institute \\
University of Minnesota \\}
\end{center}
\vspace{0.2in}
\begin{flushright}
FTPI-MINN-12/26 \\
UMN-TH-3114/12 \\
August 2012 \\
\end{flushright}
\vspace{0.3in}
\begin{center}
{\Large \bf CP Violation in Higgs Diphoton Decay in Models with Vectorlike Heavy Fermions
\\}
\vspace{0.2in}
{\bf  M.B. Voloshin  \\[0.2in] }
William I. Fine Theoretical Physics Institute, University of
Minnesota,\\ Minneapolis, MN 55455, USA, \\
School of Physics and Astronomy, University of Minnesota, Minneapolis, MN 55455, USA, \\
and \\
Institute of Theoretical and Experimental Physics, Moscow, 117218, Russia
\\[0.2in]

\end{center}

\vspace{0.2in}

\begin{abstract}
The amplitude of the diphoton decay of the Higgs boson is considered in an extension of the Standard Model involving heavy vectorlike fermions, which was recently suggested in the literature in order to explain a possible enhancement of the $h \to \gamma \gamma$ decay rate. It is pointed out that generally in such models the decay amplitude acquires  a sizable CP-odd contribution from the complex phases in the mass matrix of the new heavy fermions. The resulting CP violation in the diphoton decay can be studied experimentally when both photons are converted into $e^+ e^-$ pairs by measuring the distribution over the angle between the planes of the pairs. Such measurement, if feasible, would be of a great interest on its own, regardless specific models.
\end{abstract}
\end{titlepage}

The observed in the LHC experiments~\cite{atlas,cms} properties of the particle with mass around 126\,GeV agree with the expectations for a Higgs boson $h$ of the Standard Model (SM), except, possibly, for an aparent hint at an enhancement of the diphoton decay rate $h \to \gamma \gamma$. Namely, taken at face value, the data correspond to the two photon decay rate being larger by a factor $1.5 \div 2$ than predicted~\cite{egn,svvz} in the SM. Clearly, if such enhancement persists in the further data, this will require an extension of SM. 

The most effective way of enhancing the Higgs boson diphoton decay without significantly affecting other properties of the boson that are in agreement with the data, is identified~\cite{df,clw,alw,jsw,ahbdf,abmf,kpw, dlm} as an introduction of vectorlike heavy leptons, whose mass matrix is only partially given by Yukawa interaction with the Higgs field, and the other part is an $ad$ $hoc$  singlet with respect to the electroweak $SU(2) \times U(1)$ symmetry and is not related to a Higgs Yukawa interaction. The usage of leptons, rather than heavy quarks, allows to avoid modification of the coupling of $h$ to two gluons and thus to not affect the production of $h$ by the gluon fusion, while an introduction of vectorlike, rather than chiral, leptons breaks the known rigid proportionality between the fermion mass and the coupling to $h$. The latter property is important for the following reason. The coupling of the Higgs boson to two photons arises from loops with charged particles, and the largest contribution is given by the loop with $W$ bosons. The contribution of a charged fermion with a mass proportional to the Higgs field is of the opposite sign and is several time smaller~\cite{svvz}. Thus adding new generations of chiral fermions would only reduce the $h \to \gamma \gamma$ coupling (unless an unreasonably large number of such leptons is introduced in the model). In the case of heavy vectorlike fermions,  $SU(2) \times U(1)$ singlet mass terms can be introduced in addition to the Higgs - generated, so that with appropriate choice of the sign of the Yukawa couplings, the contribution of the loop with new leptons can be made of the same sign as the $W$ loop, thus enhancing the $h \to \gamma \gamma$ decay. 

It can be noted however, that the Yukawa sector in the discussed models of vectorlike leptons generally contains complex phases that cannot be rotated away~\cite{ahbdf}. These phases introduce violation of the CP symmetry in the Yukawa sector. The purpose of this paper is to point out that a CP violation in the mass sector results in a CP-violating contribution to the amplitude of the decay $h \to \gamma \gamma$, which can in principle be measured experimentally by studying the distribution of the angle between linear polarizations of the two photons.

The new hypothetical leptons are assumed to be heavy, so that the discussed Higgs boson is well below the threshold for pair production of the new particles. In this situation one can consider  the $h \gamma \gamma$ coupling generated by the loops with the heavy leptons, using the approach of the low energy theorems~\cite{svvz}, i.e. by calculating the loops for the two-photon correlator in a static Higgs field $\phi=v +h$, and then identifying the term linear in the excitation $h$ over the vacuum average $v = (G_F \, \sqrt{2})^{-1/2} \approx 246$\,GeV. The main observation of the present paper can be formulated as follows. Let ${\cal M}(\phi)$ be the mass matrix for the heavy fermions with charge $Q_e$, i.e. the mass term for these fermions is written as
\be
{\cal L}_M=- {\cal M}_{i,j} (v) \, ({\overline \psi}_{i L} \psi_{j R}) + h.c.~,
\label{mt}
\ee
where the indices $i$ and $j$ label the `flavors' of the heavy fermions and thus the number of these `flavors' determines the dimension of the matrix ${\cal M}$. The contribution of the heavy fermions to the $h \gamma \gamma$ coupling can then be written as the effective Lagrangian
\bea
&&{\cal L}_{h \gamma \gamma} = 
\label{hgg} \\ \nonumber 
&& \!\!\!\!\!\!\!\!\!\!{\alpha \over 4 \pi} \, Q_e^2 \,{h \over v}  \left \{ {1 \over 3} \, F^{\mu \nu} F_{\mu \nu} \, {\partial \over \partial \log v} \log {\rm Det} \left [{\cal M}^\dagger(v) {\cal M}(v) \right ]  + {1 \over 2}\, \, \epsilon^{\mu \nu \lambda \sigma}  F_{\mu \nu} F_{\lambda \sigma}  {\partial \over \partial \log v} \arg \left [{\rm Det} {\cal M}(v) \right ] \right \}~.
\eea
The CP-even term in this formula, proportional to $F^{\mu \nu} F_{\mu \nu}$, is the result of the low energy theorems~\cite{svvz} expressed in the generalized matrix form~\cite{clw}, while the CP-odd term, proportional to $F \tilde F$ and omitted in the previous studies, does not arise in SM with sequential chiral generations of fermions, and is specific for the extended models with vectorlike fermions. The latter term corresponds to the CP-odd part of the effective QED Lagrangian that one finds upon integrating out heavy fermions:
\be
\Delta{\cal L}_{\rm QED} =  {\alpha \over 4 \pi} \, Q_e^2 \, \left \{ {1 \over 3} \, F^{\mu \nu} F_{\mu \nu} \,  \log {\rm Det} \left [{{\cal M}^\dagger(v) {\cal M}(v) \over \Lambda^2} \right ]  + {1 \over 2}\, \, \epsilon^{\mu \nu \lambda \sigma}  F_{\mu \nu} F_{\lambda \sigma}   \arg \left [{\rm Det} {\cal M}(v) \right ] \right \}~ 
\label{effl}
\ee
with $\Lambda$ being an ultraviolet cutoff parameter.

As is well known, the CP-odd `anomalous' term in Eq.(\ref{effl}) by itself is of no consequence in QED since $F \tilde F$ is a total derivative. Moreover, an arbitrary constant, `the $\theta$ term', can be added to the coefficient of this part of the QED Lagrangian with no modification of physically measurable amplitudes. However, the contribution in Eq.(\ref{hgg}) of the variation of this part of the effective Largrangian in the Higgs background is of a potential physical significance, and, as discussed further, it gives rise to a measurable effect of CP violation in the decay $h \to \gamma \gamma$. One can also readily notice that in SM with only chiral fermions the CP-odd term in Eq.(\ref{hgg}) vanishes, since the mass matrix is proportional to the Higgs field, so that the complex phase of the determinant does not depend on $v$.

A nonvanishing CP-odd term in Eq.(\ref{hgg}) is naturally generated in a minimal model of vectorlike leptons. This model contains two lepton doublets of opposite chirality: $L = ( N^0, E^-)_L$ and $R= ( N^0, E^-)_R$, and two opposite chirality singlets: $S^-_L$ and $S^-_R$. The mass term for these heavy vectorlike leptons has the general form
\be
{\cal L}_{M2}=-m_1 \, ({\overline S}_L S_R) - m_2 \, ({\overline L}R) - \sqrt{2} \, y_{12}\,  ({\overline S}_L\, H^\dagger \, R)-  \sqrt{2} \, y_{21}\,({\overline L} \, H \, S_R) + h.c.~,
\label{lm2}
\ee
where $H$ is the Higgs field doublet and $y_{12}$ and $y_{21}$ are dimensionless Yukawa couplings. This model describes one neutral Dirac lepton $N^0$ with mass $m_2$ and two Dirac leptons with the electric charge $Q_e=-1$, whose $2 \times 2$~mass matrix in the basis $(S^-,\,E^-)$ has the form
\be
{\cal M}(v) =
 \left ( \begin{array}{cc}
 m_1 & y_{12} v \\
 y_{21} v & m_2 \end{array} \right )~.
\label{mm2}
\ee
For complex Yukawa couplings $y$, the phase of the determinant of this matrix, Det${\cal M}(v)=m_1 m_2 - y_{12} y_{21} v^2$, generally cannot be rotated away~\cite{ahbdf}. Moreover, there is no apriori reason to expect this phase to be small.  One can readily write explicit expressions for the contribution of the vectorlike leptons to the coefficients in Eq.(\ref{hgg}):
\be
{\partial \over \partial \log v} \log {\rm Det} \left [{\cal M}^\dagger(v) {\cal M}(v) \right ] = {-4  v^2 {\rm Re}(m_1  m_2 y_{12} y_{21}) + 4 v^4 |y_{12} y_{21}|^2 \over |m_1 m_2 - y_{12} y_{21} v^2|^2}
\label{dre}
\ee
and
\be 
{\partial \over \partial \log v} \arg \left [{\rm Det} {\cal M}(v) \right ] = -{2 v^2 {\rm Im}(m_1 m_2 y_{12} y_{21}) \over  |m_1 m_2 - y_{12} y_{21} v^2|^2}~.
\label{dim}
\ee
The phase of the product $m_1  m_2 y_{12} y_{21}$ is invariant under phase rotations of the fields of fermions and thus cannot be eliminated~\cite{ahbdf}. Furthermore, for a constructive interference with the SM amplitude of $h \to \gamma \gamma$ the expression in Eq.(\ref{dre}) has to be negative and not small, so that in particular the product $-4  v^2 {\rm Re}(m_1  m_2 y_{12} y_{21})$ should be substantially larger than $4 v^4 |y_{12} y_{21}|^2$. Therefore there is no apparent reason to expect that the coefficient (\ref{dim}) for the CP-odd part of the amplitude is suppressed relative to the coefficient (\ref{dre}) for the CP-even contribution of the vectorlike leptons.  

The total effective Lagrangian for the $h  \gamma \gamma$ interaction can thus be written in the general form
\be
{\cal L}_{h \gamma \gamma}= (A_{SM}+A_{VL}^+) \, h \, F^{\mu \nu} F_{\mu \nu} + A_{VL}^- \, h \, {1 \over 2} \, \epsilon^{\mu \nu \lambda \sigma} F_{\mu \nu} F_{\lambda \sigma}
\label{lphgg}
\ee
with $A_{SM}$ being the SM contribution and $A_{VL}$ being the contribution from the vectorlike leptons with $A_{VL}^+$ ($A_{VL}^-$) standing for its CP-even (CP-odd) part. 
Introducing the unit vector $\vec n$ along the direction of the final photons in the rest frame of the decaying Higgs boson and also the transverse vectors $\vec a_1$ and $\vec a_2$ for the linear polarization of the photons [$(\vec a_1 \cdot \vec n)= (\vec a_2 \cdot \vec n)=0$], one can find, from Eq.(\ref{lphgg}) the amplitude for the decay $h \to \gamma \gamma$ in this reference frame as
\be
A(h \to \gamma \gamma)= - 2 \, M_H^2 \, \left \{ (A_{SM}+A_{VL}^+) \, (\vec a_1 \cdot \vec a_2) +   A_{VL}^- \, [\vec n \cdot (\vec a_1 \times \vec a_2)] \right \}~.
\label{ahgg}
\ee 
The amplitudes of different CP parity do not interfere in the total decay rate, $\Gamma(h \to \gamma \gamma) \propto (A_{SM}+A_{VL}^+)^2 +  (A_{VL}^-)^2$. They however do interfere in the distribution over the angle $\vartheta$ between the linear polarizations of the two photons. The decay amplitude as a function of this angle reads as
\be
A(h \to \gamma \gamma) \propto (A_{SM}+A_{VL}^+)\, \cos \vartheta +  A_{VL}^- \sin \vartheta \propto
\cos (\vartheta - \beta)
\label{ang}
\ee
with the CP-violation effect encoded in the angle $\beta$ as
\be
\tan \beta = {  A_{VL}^-  \over A_{SM}+A_{VL}^+}~.
\label{cpb}
\ee
The distribution of the decay rate over the angle $\theta$ is given by
\be
{d \Gamma(h \to \gamma \gamma) \over d \vartheta}  \propto \cos^2 (\vartheta - \beta)~,
\label{ad}
\ee 
so that the maximum of the distribution is at $\vartheta = \beta$.
For the SM Higgs boson any nonzero value of $\beta$ would imply violation of the CP symmetry in the diphoton decay.~\footnote{It can be also mentioned that if the observed 126\,GeV particle is not the SM Higgs boson, a value of $\beta = \pi/2$ would unambiguously identify this particle as a pseudoscalar, rather than imply a CP violation.}

In order to explain an enhancement by a factor $1.5 \div 2$ of the diphoton decay rate, one should assume that  $A_{VL}^+/A_{SM} \approx 0.2 \div 0.4$. Since, as discussed, one can generally expect that the CP-odd amplitude $A_{VL}^-$ is not suppressed relative to the CP-even  $A_{VL}^+$, the angle $\beta$ can easily amount to few tenths of a radian, or more. 

The discussed angular distribution can be experimentally measured when both photons are converted, either internally, or in the detector, into $e^+e^-$ pairs  by the angular distribution between the planes defined by the momenta of the $e^+$ and $e^-$ in the pairs (in essentially similar way as has been done for the $\pi^0$ decay~\cite{planoea}). It would be extremely interesting if such measurement could be done with the current LHC detectors, or in the future.~\footnote{The CP symmetry in the $h \to \gamma \gamma$ interaction can also be studied in the Higgs boson production at $\gamma \gamma$ and $e \gamma$ colliders~\cite{gi}.} Clearly, such test of CP symmetry in the decay $h \to \gamma \gamma$ may provide an important insight, even if specific models of heavy vectorlike fermions turn out to be irrelevant for the actual nature. 

I thank Yuri Gershtein for discussion of feasibility of detailed experimental study of the $h \to \gamma \gamma$ decay.
This work is supported, in part, by the DOE grant DE-FG02-94ER40823. I acknowledge the hospitality 
of the Aspen Center for Physics, where this paper was written during the program supported in part by the National Science Foundation under Grant No. PHY-1066293.

\end{document}